\documentclass[sigconf]{acmart}

\AtBeginDocument{%
  \providecommand\BibTeX{{%
    \normalfont B\kern-0.5em{\scshape i\kern-0.25em b}\kern-0.8em\TeX}}}

\copyrightyear{2022}
\acmYear{2022}
\setcopyright{acmcopyright}
\acmConference[MM '22] {Proceedings of the 30th ACM International Conference on Multimedia }{October 10--14, 2022}{Lisbon, Portugal.}
\acmBooktitle{Proceedings of the 30th ACM International Conference on Multimedia (MM '22), October 10--14, 2022, Lisbon, Portugal}
\acmPrice{15.00}
\acmISBN{978-1-4503-9203-7/22/10}
\acmDOI{10.1145/3503161.3548193}

\usepackage{multirow, booktabs}
\usepackage{colortbl}
\usepackage{color,xcolor}
\usepackage{booktabs}
\usepackage{threeparttable}
\usepackage{titlesec}
\usepackage{url}
\usepackage{float}
\usepackage{amsmath}

\begin{document}

\title{DeepWSD: Projecting Degradations in Perceptual Space to Wasserstein Distance in Deep Feature Space}

\author{Xingran Liao}
\affiliation{%
  \institution{City University of Hong Kong}
  \city{Hong Kong}
  \country{China}}
\email{xingrliao2-c@my.cityu.edu.hk}

\author{Baoliang Chen}
\affiliation{%
  \institution{City University of Hong Kong}
  \city{Hong Kong}
  \country{China}}
\email{blchen6-c@my.cityu.edu.hk}

\author{Hanwei Zhu}
\affiliation{%
  \institution{City University of Hong Kong}
  \city{Hong Kong}
  \country{China}}
\email{hanwei.zhu@my.cityu.edu.hk}

\author{Shiqi Wang}
\affiliation{%
  \institution{City University of Hong Kong}
  \city{Hong Kong}
  \country{China}}
  \email{shiqwang@cityu.edu.hk}
  
\author{Mingliang Zhou}
\affiliation{%
  \institution{Chongqing University}
  \streetaddress{174 shazheng street, Shapingba District}
  \city{Chongqing}
  \country{China} }
\email{mingliangzhou@cqu.edu.cn}
  
\author{Sam KWONG}
\affiliation{%
  \institution{City University of Hong Kong}
  \streetaddress{83 Tat Chee Ave, Kowloon Tong}
  \city{Hong Kong}
  \country{China}
  \postcode{999077} }
\email{cssamk@cityu.edu.hk}

\renewcommand{\shortauthors}{Xingran Liao et al.}

\begin{abstract}
Existing deep learning-based full-reference IQA (FR-IQA) models usually predict the image quality in a deterministic way by explicitly comparing the features, gauging how severely distorted an image is by how far the corresponding feature lies from the space of the reference images. Herein, we look at this problem from a different viewpoint and propose to model the quality degradation in perceptual space from a statistical distribution perspective. As such, the quality is measured based upon the Wasserstein distance in the deep feature domain. More specifically, the 1D Wasserstein distance at each stage of the pre-trained VGG network is measured, based on which the final quality score is performed. {The deep Wasserstein distance (DeepWSD) performed on features from neural networks enjoys better interpretability of the quality contamination caused by various types of distortions and presents an advanced quality prediction capability.} Extensive experiments and theoretical analysis show the superiority of the proposed DeepWSD in terms of both quality prediction and optimization. The implementation of our method is publicly available at \url{https://github.com/Buka-Xing/DeepWSD}.
\end{abstract}



\maketitle

\section{Introduction}

Full-reference image quality assessment (FR-IQA), which establishes the quantitative projection between perceptual quality and computational models~\cite{zhoubook}, plays an essential role in monitoring the quality of multimedia processing systems. In addition, the FR-IQA models have been shown to excel in quality optimization of image restoration tasks, such as denoising, deblurring and superresolution~\cite{IQAopt,ADISTS}. 

The design philosophies of the FR-IQA models are supported by the statistics of natural scenes, the prior knowledge of the Human Visual System (HVS), as well as the understanding of distortion~\cite{BayesIQA}. In particular, it is widely accepted that extraction of the features that account for naturalness from reference and distortion images is important. The well-known structural similarity (SSIM) index~\cite{SSIM} focuses on the structural information from a scene. The deep image structure and texture similarity (DISTS) method~\cite{DISTS} adopts the pretrained deep convolutional network (VGG16) learned from abundant images to extract features. The FR-IQA models also attempt to mimic the visual perception in HVS. The feature similarity index (FSIM)~\cite{FSIM} adopts the log-Gabor filters to mimic such mechanism, supported by the psychophysical evidence~\cite{Log-Gabor1, Log-Gabor2} that log-Gabor filters are highly correlated with the shallow visual system. The complex wavelet structural similarity (CW-SSIM) index~\cite{CWSSIM} performs the wavelet transformation to mimic the visual system, which is also supported by neurological experiments~\cite{wavelet1}. Moreover, the FR-IQA methods also consider the distortion types and levels between the reference and distortion images. In image information and visual quality (VIF)~\cite{VIF} and the information fidelity criterion (IFC)~\cite{IFC}, distortions are treated as information loss and images are constructed as Gaussian scale mixture models. Then, mutual information is used to quantify the information loss. In the most apparent distortion metric (MAD)~\cite{MAD}, distortions are classified as apparent and invisible distortions. As such, both the pyramid structure and log-Gabor filters are used to identify the distortion type and strength.

The FR-IQA shall be regarded as the fidelity measure in a wider sense, due to the existence of the reference image. Typically, the common pipeline is comparing the prominent visual information that governs the visual quality from pristine and distorted images, in spatial, frequency, or deep feature domains. These mathematical models, which are expected to model the perceptual functionalities in assessing the visual quality, typically involve making comparisons. As such, a typical deterministic way is explicitly performing feature level comparison, where the features are defined in a wider sense. However, the nonlinearities of the complicated perceptual system, as well as the capability of quality assessment without the pristine reference for HVS, persuade us to re-examine the quality assessment in a statistical way. In particular, we propose the deep feature domain Wasserstein distance (DeepWSD) for FR-IQA. The proposed method attempts to relate the FR-IQA to the minimal efforts that account for the conversion between the distributions of the pristine and distorted images in the deep feature domain. The amount of minimal effort is quantified based upon the concept of the Wasserstein distance (WSD)~\cite{OTbook}, a widely used statistical measure between two distributions. 

The proposed DeepWSD is intrinsically aligned with the efficient coding hypothesis that the efficient coding neural system is ought to match the statistics of the signals that they attempt to represent~\cite{Efficient_Coding}, from the perspective of natural scene statistics. Regarding the human perception, the free-energy principle~\cite{friston2006free, friston2010free, gu2014using, zhai2011psychovisual} also reveals that the internal generative model with which the human vision tries to explain the perceived scene, is able to produce the intrinsic reference for statistical discrepancy evaluation. The visual quality, which is governed by this discrepancy, is characterized by the Wasserstein distance here. Moreover, our method does not rely on training with quality labels or any controllable parameters for tuning, such that it is highly robust to diverse types of distortions. Experimental results show that the proposed DeepWSD delivers accurate predictions over a series of datasets, and enjoys the advantage of strong generalization capability and high interpretability.

\section{Related Works}
\subsection{FR-IQA with Feature Comparisons}
One widely accepted philosophy in FR-IQA design is comparing perceptually meaningful visual features. From the statistical perspective, image features can be treated as invariants of the distribution~\cite{Invariants} which contain important information of images. Distortions typically occur~\cite{Invariants3} along with the contamination of these invariants. Traditional methods extract image features in the pixel or the frequency domain, based upon the similarity measure or the Euclidean norm. Regarding SSIM~\cite{SSIM}, the luminance, contrast, and structure are treated as features in a wider sense, and they are compared by the similarity measure. 

In FSIM~\cite{FSIM}, the high phase congruent features in the frequency domain are extracted by the log-Gabor filters and features are also compared by the similarity index. FSIM~\cite{FSIM} presents outstanding performance with increased computation cost. In Gradient Magnitude Similarity Deviation (GMSD)~\cite{GMSD}, image gradients are extracted by Prewitt operators and compared by simply using the Euclidean norm, due to the observation that gradient map of images is more sensitive to various types of distortions~\cite{GMSD}. 

Due to the excellent performance of deep neural networks in feature representation, deep features can also be compared to indicate the visual quality. In Learned Perceptual Image Patch Similarity (LPIPS)~\cite{LPIPS}, different deep network encoders have been used to extract different deep features. Subsequently, LPIPS adopts weighted MSE for comparisons, presenting better performance compared to most traditional FR-IQA methods~\cite{LPIPS}. In DISTS~\cite{DISTS}, motivated by the design of the modified SSIM~\cite{addSSIM}, deep features are compared in a new way, delivering more accurate subjective ratings. Based upon DISTS~\cite{DISTS}, the A-DISTS~\cite{ADISTS} which separates the structure and texture across space and scale, further improves the performance of DISTS~\cite{DISTS}. However, the explicit comparisons of the features in spatial, frequency, or deep learning domains do not specifically consider the image quality from the statistical perspective, while a large literature has been dedicated to the importance of naturalness from the statistical point of view in biological vision~\cite{NSS}. 

\subsection{Wasserstein Distance}  

The Wasserstein distance originates from optimal transformation~\cite{OTbook} and has been widely used in different vision tasks. In particular, it characterizes the differences between two distributions by computing how difficult a specific distribution can be transformed into another. In image retrieval~\cite{WSD_imageretrival}, the Wasserstein distance was used to compare the similarity between the histogram of color, texture, and outlines of images. It delivers higher retrieval accuracy than traditional Euclidean norm and Chi-square distance~\cite{chi_square}. In image reconstruction and generation tasks~\cite{WSD_denoise,WGAN}, the Wasserstein distance also exhibits promising capability on texture generation compared to the Euclidean norm and the Kullback-Leibler divergence~\cite{KLD}. Moreover, using the Wasserstein distance in Generative-Adversarial networks can better prevent mode collapse than using the Kullback-Leibler divergence~\cite{WGAN}. In image style transfer ~\cite{WSD_style}, the distillation method based on Wasserstein distance also exhibits that such distance can better maximize the lower bound of mutual information between source and target images compared to the Kullback-Leibler divergence, leading to the reduction of artifacts. However, to the best of our knowledge, the Wasserstein distance has not been exploited in the FR-IQA. 

\section{Quality Assessment Method}

\subsection{Methodology}
In DeepWSD, we evaluate the quality of distorted images by using 1D Wasserstein distance~\cite{WSD}. In particular, we first adopt the VGG16 network~\cite{VGG16} which projects images into deep feature representations, and subsequently, the 1D WSD is used to compare the distributions of the input images as well as all the deep features. In the following paragraphs, we first provide a brief introduction of WSD, then the proposed method is detailed.

\subsubsection{Preliminary of WSD}
Given two multidimensional random variables $M$ and $N$ with their distributions denoted as $\mathcal{X}$ and $\mathcal{Y}$, respectively, the Wasserstein distance is defined as,
\begin{equation}
W_{l}(M,N)={{\left(\underset{J\in \mathcal{J}(\mathcal{X},\mathcal{Y})}{\mathop{\inf }}\,\int{||m-n|{{|}_{l}}\;dJ(m,n)}\right)}^{{1}/{l}\;}},
\label{Eq.WSD_define}
\end{equation}
where $m$ and $n$ are the masses of $M$ and $N$. J $\in$ $\mathcal{J}(\mathcal{X},\mathcal{Y})$ is the joint distribution of $(M,N)$. $l$ is the order of the $l$-norm. Typically, $l$ can be set as 1 or 2, and in our experiments we set $l=2$. Supposing  $F(m)=P\{M\leqslant m\}$ and $G(n)=P\{N\leqslant n\}$, as indicated in~\cite{OTbook}, when  $M$ and $N$ are one-dimension, Eqn. (\ref{Eq.WSD_define}) has a closed-form solution. It computes the integral of the inverse function of the probability distribution function of $\mathcal{X}$ and $\mathcal{Y}$,
\begin{equation}
W_{l}(M,N)={{\left(\int_{0}^{1}{||{{F}^{-1}}(z)-{{G}^{-1}}(z)||_l \; dz}\right)}^{{1}/{l}\;}},
\label{Eq.1D-WSD}
\end{equation}
where $z$ is the implicit variable which is used to integral $F^{-1}(\cdot)$ and $G^{-1}(\cdot)$ from 0 to 1.  $F^{-1}(\cdot)$ and $G^{-1}(\cdot)$ represent the inverse functions of $F(\cdot)$ and $G(\cdot)$, respectively. 

\subsubsection{The Proposed DeepWSD}
\begin{figure}[t]
\centering
      \includegraphics[scale=0.30]{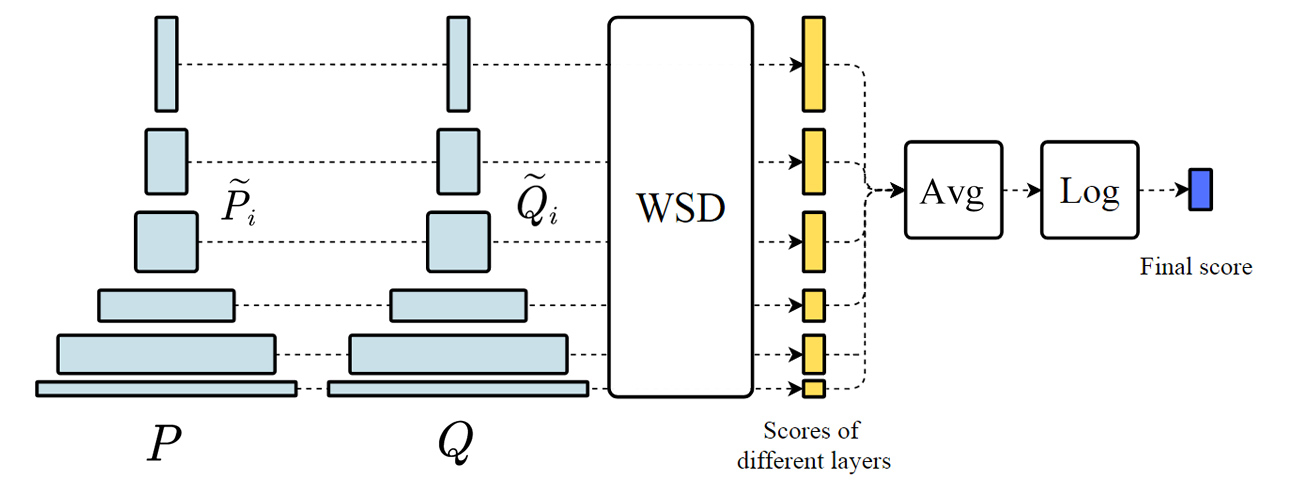}
      \caption{Overall structure of DeepWSD. It contains  the raw image and five feature extraction stages in VGG16. `Avg' means averaging scores in different layers and `Log' indicates the logarithm function.}
      \label{fig:DeepWSD}
\end{figure}

In DeepWSD, we adopt the one-dimension form of WSD for the distribution divergence measuring. More specifically, as shown in Fig.~\ref{fig:DeepWSD}, the input reference image and distorted image are denoted as  $P$ and $Q$, respectively. Then we utilize the pre-trained VGG16 network~\cite{VGG16} to extract their deep features at five stages. In particular, we denote the extracted feature of  $P$ and $Q$ at the $i-th$ stage as $\tilde{P}_i$ and  $\tilde{Q}_i$, $i \in \{1,5\} $. Herein, the VGG16 network was originally used to deal with the image classification task and has been widely used in different CV tasks including super-resolution~\cite{VGG_SR} and denoising~\cite{VGG_denoise}. In LPIPS~\cite{LPIPS} and DISTS~\cite{DISTS}, features of the VGG16 network were also proved to be useful for quality assessment.  Subsequently, we reshape the $P$, $Q$,  $\tilde{P}_i$, and  $\tilde{Q}_i$ to 1D vectors thus the WSD can be performed as follows, 
\begin{equation}
\mathcal{D}_{wsd}\left( P,Q \right) = W_l\left(P,Q \right) + \sum_{i=1}^5{W_l\left( \tilde{P}_i,\tilde{Q}_i \right)}.
\label{Eq.DeepWSD}
\end{equation}
In addition to WSD, we further employ the adaptive Euclidean distance (denoted as $\mathcal{D}_{eul}$) to pursue high fidelity, which can be formulated as follows,
\begin{equation}
\begin{aligned}
\mathcal{D}_{eul}\left( P,Q \right) = & g(W_l\left(P,Q \right))\times ||P - Q||_2 \\
&+ \sum_{i=1}^5{\left(g\left(W_l\left( \tilde{P}_i,\tilde{Q}_i\right)\right)\times ||\tilde{P}_i - \tilde{Q}_i||_2 \right)}  ,
\label{Eq.DeepWSD}
\end{aligned}
\end{equation}
where the definition of $g(s)$ is given by,
\begin{equation}
g(s) = \frac{1}{{{(s+10)}^{2}}\sqrt{\exp (-{1}/{(s+10)}\;)}}.
\label{Eq.g(s)}
\end{equation}
The combination of Euclidean norm and adaptive weight $g(s)$ mimics a special mechanism when HVS quantifies the visual quality. In particular, Larson \textit{et al.}~\cite{MAD} pointed out that the most apparent distortions are first perceived in a high priority. Then, scrupulous viewers may find certain parts of the images contain subtle distortions.  By considering the $\mathcal{D}_{wsd}$  and $\mathcal{D}_{eul}$ simultaneously, the distance scores of each layer are finally averaged and further processed by a logarithmic function (denoted as $\log(\cdot)$), deriving the proposed DeepWSD measure as follows,

\begin{equation}
DeepWSD \left( P,Q \right) = \log \left(\frac{1}{6} \mathcal{D}_{wsd}\left( P,Q \right)+\frac{1}{6}  \mathcal{D}_{eul}\left( P,Q \right)\right).
\label{Eq.DeepWSD2}
\end{equation}
As a deep learning based quality measure, the DeepWSD avoids the training with quality labels, enjoying the great advantages in terms of generalization capability. Moreover, the proposed DeepWSD does not rely on any hyperparameter that is determined empirically regarding the generation of the final score. We believe that these are the directions in which the greatest benefit eventually lies.

\begin{figure}[tbp]
\centering
      \includegraphics[scale=0.35]{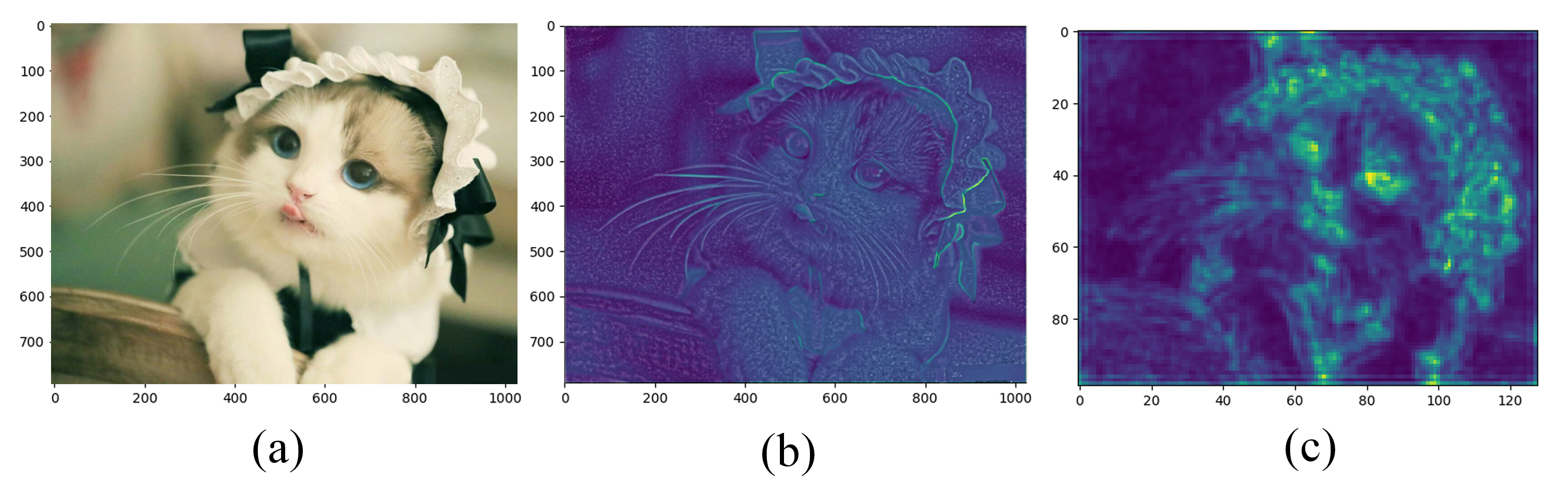}
      \caption{(a) original image; (b) structure map of (a) extracted by SSIM; (c) feature map sampled in the 4-th layer of VGG16.}
      \label{fig:cat}
\end{figure}

\begin{figure*}[tbp]
\centering
      \includegraphics[scale=0.50]{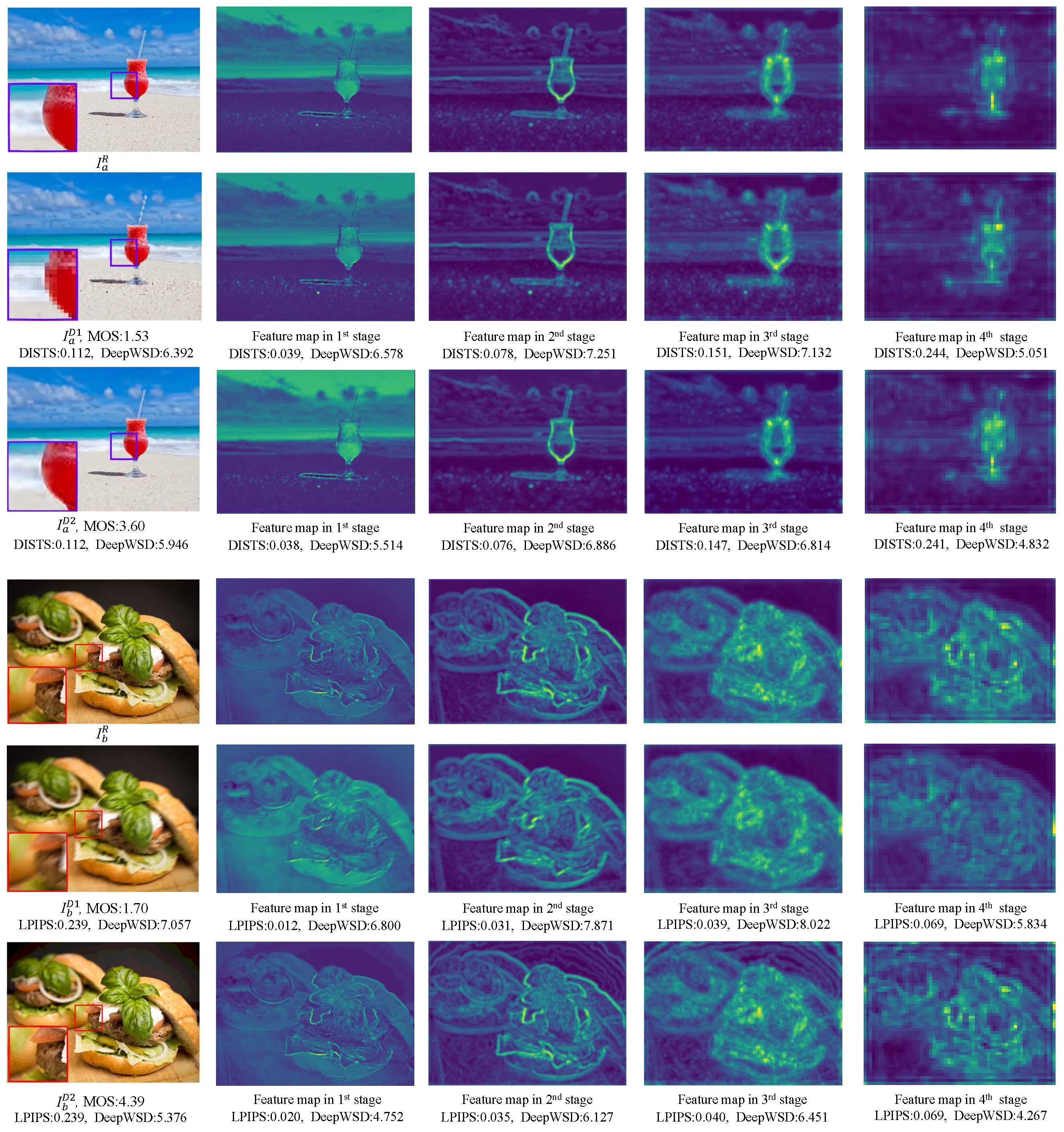}
      \caption{Quality measure results in feature domain by different quality models. $I_{a}^{R}$ and $I_{b}^{R}$ are two reference images. $I_{a}^{D1}$  ($I_{b}^{D1}$ ) and  $I_{a}^{D2}$ ($I_{b}^{D2}$) are two  distorted versions of  $I_{a}^{R}$  ($I_{b}^{R}$ ) with different quality ratings. The scores under each distorted image are the final outputs of corresponding IQA models and the scores of the feature maps are calculated by the feature comparison method adopted in each IQA model. 
      }
      \label{fig:ftshow}
\end{figure*}

\renewcommand{\arraystretch}{1.1}
\begin{table*}[tp]
  \centering
  \fontsize{10}{10}\selectfont
  \begin{threeparttable}
  \caption{Comparison between our method and state-of-the-art methods on four classical datasets. Bold and uderline numbers correspond to the first and second best methods in each column, respectively.}
  \label{table:classic}
    \begin{tabular}{ccccccccccccc}
    \toprule[2pt] \midrule[1pt]
    \multirow{2}{*}{Method}&
    \multicolumn{3}{c}{TID2013~\cite{TID2013}} & \multicolumn{3}{c}{LIVE~\cite{LIVE}} & \multicolumn{3}{c}{CSIQ~\cite{MAD}} & \multicolumn{3}{c}{IVC~\cite{IVC}} \cr
    \cmidrule(lr){2-4} \cmidrule(lr){5-7} \cmidrule(lr){8-10}  \cmidrule(lr){11-13} 
    &PLCC&SRCC&KRCC&PLCC&SRCC&KRCC&PLCC&SRCC&KRCC&PLCC&SRCC&KRCC \cr
    \hline
    PSNR &0.6773 &0.6876 &0.4963 &0.8648 &0.8726 &0.6773 &0.8195 &0.8101 &0.6014 &0.6804 &0.6499 &0.4781 

 \\
    SSIM~\cite{SSIM} &0.7767 &0.7271 &0.5454 &0.9341 &0.9479 &0.7963 &0.8523 &0.8656 &0.6807 &0.7708 &0.9018 &0.7223 
 \\
    
    MS-SSIM~\cite{MS-SSIM} &0.8292 &0.7887 &0.6049 &0.9399 &0.9513 &0.8048 &0.8895 &0.9133 &0.7393 &0.7913 &0.8980 &0.7203 
 \\
    NLPD~\cite{NLPD} &0.8398 &{0.8491} &{0.6785} &0.9316 &0.9372 &0.7781 &0.9234 & 0.9343 &{0.7695} &0.7815 &0.8202 &0.6093 
 \\
    
    FSIM~\cite{FSIM} &0.8233 &\underline{0.8549} &0.6549 &0.9608 &\textbf{0.9672} &\textbf{0.8814} &0.9187 &0.9379 &0.7683 & 0.8161 &\textbf{0.9263} &\underline{0.7537} 
 \\
    CW-SSIM~\cite{CWSSIM} &0.5846 &0.7560 &0.5580 &0.5346 &0.9082 &0.7739 &0.6808 &{0.9441} &0.7594 &0.5630 &0.5834 &0.4072 
 \\
    
    VIF~\cite{VIF} &\underline{0.8672} &0.8429 &0.6526 &0.9343 &0.9603 &0.8284 &0.9132 &0.9121 &0.7432 &0.7391 &0.7273 &0.5590 
 \\ 
    MAD~\cite{MAD}&0.8267 &0.7680 &0.6154 &\textbf{0.9682} & \underline{0.9669} & \underline{0.8425} & \textbf{0.9505} & \underline{0.9468}  & \underline{0.7975} & 0.8704 & 0.8698 & 0.6671 
 \\
    
    PieAPP~\cite{PieAPP}& 0.8501 &0.8479 &\underline{0.6828} &0.9079 &0.9279 &0.8266 &0.8823 &0.8833 &0.7305 &0.6594 &0.7187 & 0.5249 
 \\
    DISTS~\cite{DISTS} &{0.8624} &0.8483 &0.6574 &0.9360 & 0.9542 & 0.8112 &{0.9284} & 0.9289 & 0.7675 &\underline{0.8993} &{0.9138} &0.7267 
 \\
    
    LPIPS~\cite{LPIPS} &0.7324 & 0.6696 & 0.4970 &0.9343 &0.9324 & 0.7782 & 0.8936 & 0.8758 & 0.6893 & {0.8715} &0.9044 &{0.7386}
 \\
    DeepWSD &\textbf{0.8697} &\textbf{0.8741} &\textbf{0.6932} &\underline{0.9609} &{0.9624} &{0.8378} &\underline{0.9503} &\textbf{0.9652} &\textbf{0.8297} &\textbf{0.9119} &\underline{0.9231} &\textbf{0.7570} 
\\
    \midrule[1pt] \bottomrule[2pt]
    \end{tabular}
    \end{threeparttable}
\end{table*}

\subsection{Connections with Existing Models}

The new design philosophy of DeepWSD can be better understood by exploring the connections with existing methods. First, the traditional quality measures, based upon the point-wise signal difference, structural similarity quantification, or feature-domain comparisons, suggest that exploiting the strong dependencies in visual representation is important in modeling the visual quality. One intuitive example is shown in Fig.~\ref{fig:cat}, which suggests that both structures and deep features reflect such dependencies. The proposed DeepWSD naturally considers such correlations in the deep feature domain due to its high relevance in visual perception. 

Second, it is widely acknowledged that the features from the pre-trained VGG network correlate well with the visual perception~\cite{DISTS,LPIPS}. As such, the straightforward point-wise feature map comparisons, or conducting the structure and texture similarity modeling, could better quantify the visual quality. The proposed method adopts the deep learning features as well, while the principle philosophy underlying the companions has been advanced to a new level with the probability modeling on the distribution. As shown in Fig.~\ref{fig:ftshow}, the distortions introduced cause the significant quality degradation in the feature domain, demonstrating that deep features possess a strong capability for sensing distortions. Moreover, we can observe that even the two distorted images, $I_{a}^{D1}$ and  $I_{a}^{D2}$ present totally different quality (1.53 \textit{v.s.} 3.60), the feature comparison results obtained by DISTS are very close at each stage. On the contrary, the quality measured by our DeepWSD could better reflect the distortions and such variations. 
The quality unawareness is also presented by LPIPS in Fig.~\ref{fig:ftshow} when compared with DeepWSD, further revealing the effectiveness of WSD performed on deep feature space. In ~\cite{WSD_ImageEnhance}, the optimization of the WSD of deep features has been shown to be effective in low-level visual processing, providing further evidence that the exploration of the WSD for quality assessment is meaningful. 

Third, the assumption that HVS is highly adapted to the natural scene has been a golden hypothesis in no-reference IQA. Due to the unavailability of the pristine images, the visual quality is quantified from the perspectives that how those regularities are modified along with the distortion~\cite{chen2021no}. The FR-IQA methods, which shall be understood as the fidelity measurement, typically do not rely on distributions. The proposed approach, although belonging to FR-IQA, fundamentally lies between the evaluation of fidelity and quality, which are sometimes driven by competing factors~\cite{Tone-Map}. The visual quality is highly relevant to naturalness which has been long recognized as the subject quantity relying on the statistics of natural scene distribution. As such, the visual quality evaluation relying on the distribution of features could reliably capture low-level statistics by measuring the destruction against the feature distribution of pristine images.

\section{Experimental Results}
We evaluate the performance of DeepWSD on six IQA datasets, including  CSIQ~\cite{MAD}, TID2013~\cite{TID2013}, LIVE~\cite{LIVE}, IVC~\cite{IVC}, KADID-10k~\cite{KADID-10k} and LIVE-MultiDist~\cite{LIVE-MultiDist}. Those databases contain various distortion types and distortion levels. In particular, the KADID-10k~\cite{KADID-10k} dataset contains 81 original reference images and 25 distortion types, resulting in a total number of 10,125 images. In  LIVE-MultiDist~\cite{LIVE-MultiDist}, two mixed distortion types are involved: blur with Gaussian white noise and blur with JPEG compression, which target at the application of image acquisition and transmission. We compare DeepWSD with 11 state-of-the-art FR-IQAs, including PSNR, SSIM~\cite{SSIM}, MS-SSIM~\cite{MS-SSIM}, NLPD~\cite{NLPD}, FSIM~\cite{FSIM}, CW-SSIM~\cite{CWSSIM}, VIF~\cite{VIF}, MAD~\cite{MAD}, perceptual image-error assessment through pairwise preference (PieAPP)~\cite{PieAPP}, DISTS~\cite{DISTS} and LPIPS (with VGG16 utilized as the feature extractor)~\cite{LPIPS}. Among them, PSNR, SSIM~\cite{SSIM}, MS-SSIM~\cite{MS-SSIM} and NLPD~\cite{NLPD} are FR-IQAs computed in the pixel domain. FSIM~\cite{FSIM}, CW-SSIM~\cite{CWSSIM}, VIF~\cite{VIF} and MAD~\cite{MAD} are computed in the frequency domain. PieAPP~\cite{PieAPP}, DISTS~\cite{DISTS}, LPIPS~\cite{LPIPS} and DeepWSD are FR-IQAs computed in the deep feature domain. We use three measures to evaluate the quality assessment results, including Pearson linear correlation coefficient (PLCC), Spearman rank correlation coefficient (SRCC), and Kendall rank correlation coefficient (KRCC). In addition, a four-parameter regression model is used to visualize the score predicting results, and the definition is given by
\begin{equation}
\bar{D} = (a_1 - a_2)/(1+\exp(-(D-a_3)/|a_4|)+a_2),
\label{4_parameter}
\end{equation}
where $\bar{D}$ is the fitted score and $D$ is the raw score. $\{a_i\}_{i=1}^4$ are the parameters. For DeepWSD, we set $l=2$, making the quality measure more sensitive to outliers. For the one-dimensional case of the Wasserstein distance, we split the images into patches and reshape them as vectors. The patch size is set to $4 \times4$. A larger patch size could persuade DeepWSD to pay more attention to global differences.

\renewcommand{\arraystretch}{1.1} 
\begin{table}[tbp]
  \centering
  \fontsize{8}{8}\selectfont
  \begin{threeparttable}
  \caption{Quality prediction results on the KADID-10k and LIVE-MultiDist datasets. Bold and underline numbers correspond to the first and second best methods in each column. Herein, it is worth mentioning DISTS was trained on the KADID-10k dataset. }
  \label{table:state-of-arts}
    \begin{tabular}{ccccccc}
    \toprule[2pt] \midrule[1pt]
    \multirow{2}{*}{Method}&
    \multicolumn{3}{c}{KADID-10k~\cite{KADID-10k}} & \multicolumn{3}{c}{LIVE-MultiDist~\cite{LIVE-MultiDist}} \cr
    \cmidrule(lr){2-4} \cmidrule(lr){5-7}
    &PLCC&SRCC&KRCC&PLCC&SRCC&KRCC \cr
    \hline
    PSNR &0.6747 &0.6665 &0.4836 &0.7622 &0.7664 &0.5830 
 \\
    SSIM~\cite{SSIM} &0.7217 &0.7325 &0.5572 &0.8652 &0.8255 &0.6157 
 \\
    
    MS-SSIM~\cite{MS-SSIM} &0.8012 &0.8029 &0.6088 &0.8826 &0.8795 &0.6431 
 \\
    NLPD~\cite{NLPD} &0.8131 &0.8444 &0.6461 &0.8697 &0.8536 &0.6837 
 \\
    
    FSIM~\cite{FSIM} &\underline{0.8511} &{0.8542} &{0.6648} &0.8386 &{0.8932} & {0.6961} 
 \\
    CW-SSIM~\cite{CWSSIM} &0.5493 &0.7444 &0.5594 &0.3936 &0.4636 &0.3166 
 \\
    
    VIF~\cite{VIF} &0.7927 &0.7906 &0.6014 &{0.8947} &0.8874 &0.5428 
 \\
    MAD~\cite{MAD} &0.8232 &0.7965 &0.6210 &0.8542 &0.8712 &0.6943 
 \\
    
    PieAPP~\cite{PieAPP} &0.7889 &0.7857 &0.6493 &0.8763 &0.8881 &0.6897 
 \\
    DISTS~\cite{DISTS} &\textbf{0.8589} &\textbf{0.8893} &\textbf{0.7100} &\underline{0.8951} &\textbf{0.9330} &\underline{0.7478} 
 \\
    
    LPIPS~\cite{LPIPS} &0.7003 &0.7200 &0.5313 &0.8350 & 0.8078 & 0.6083 
 \\
    DeepWSD &{0.8434} & \underline{0.8883} & \underline{0.7076} &\textbf{0.8952} &\underline{0.9073} &\textbf{0.7558}
 \\
    \midrule[1pt] \bottomrule[2pt]
    \end{tabular}
    \end{threeparttable}
\end{table} 

\subsection{Performance Validations}
\begin{figure*}[htbp]
\centering
      \includegraphics[scale=0.50]{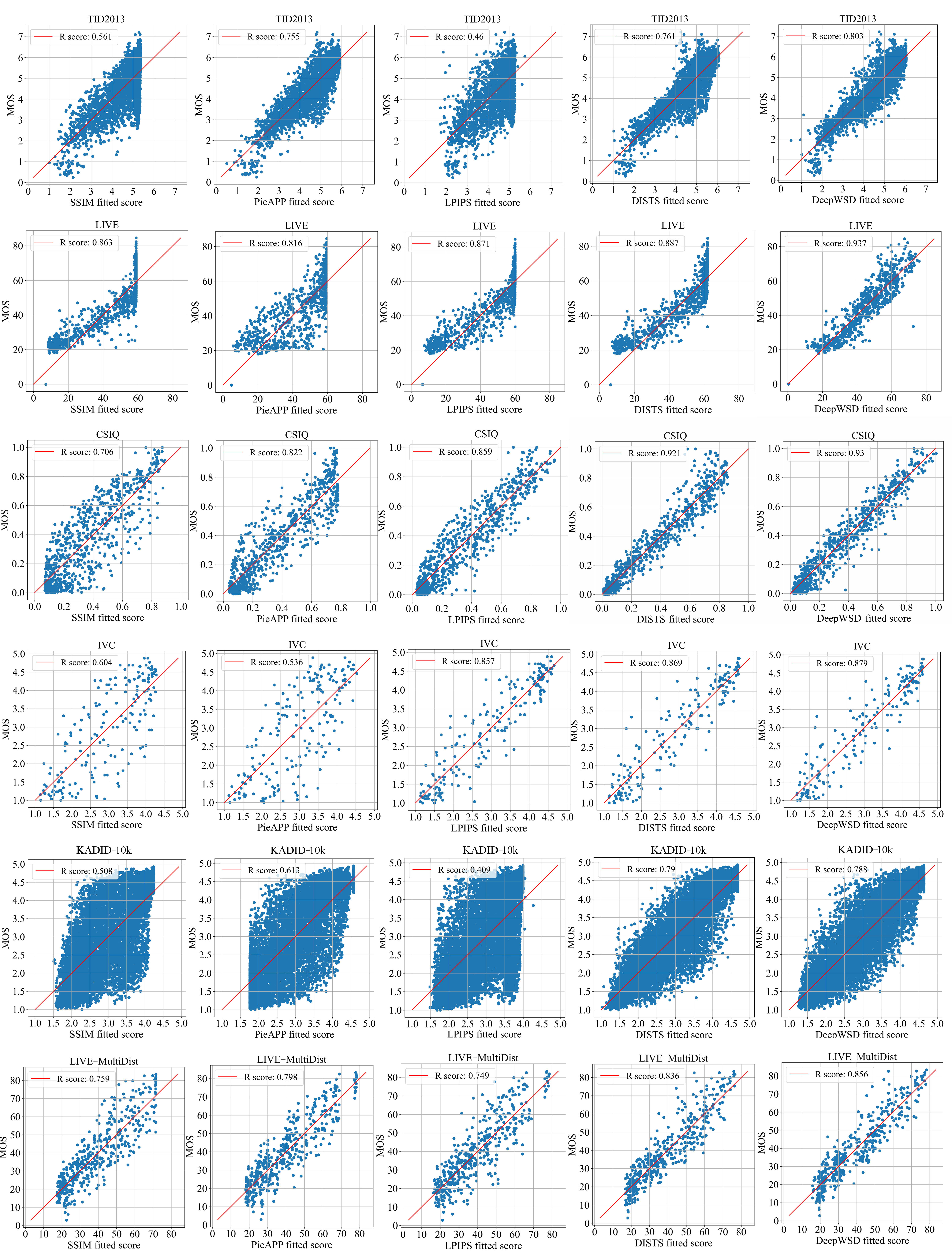}
      \caption{Visualization of regression results. The $x$-axis is the fitted scores by Eqn. (\ref{4_parameter}), and the $y$-axis is the MOS. $R=\sqrt{1-RSS/TSS}$ is the goodness of fit. Herein, $RSS$ and $TSS$ are sum of squares of deviations and total sum of squares. A larger \(R\) implies the stronger relationship between the predicted scores and MOS values. }
      \label{fig:regression}
\end{figure*}

\begin{figure*}[htbp]
\centering
      \includegraphics[scale=0.39]{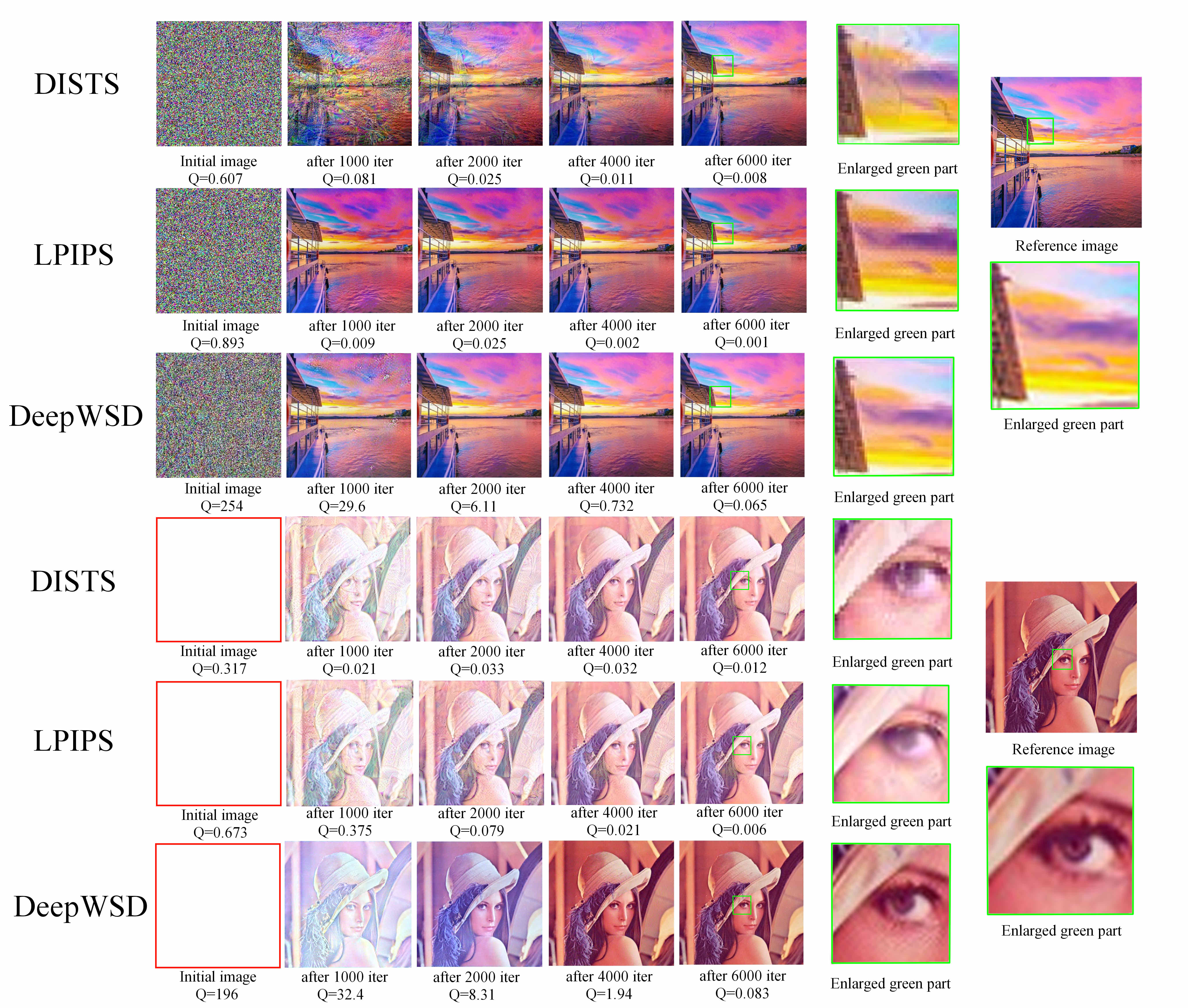}
      \caption{Optimization results by the DISTS, LPIPS, and DeepWSD. The noise-to-land and the white-to-Lena experiments are conducted. `Q' means the quality score predicted by each IQA measure.}
      \label{fig:opt}
\end{figure*}

\textbf{\textit{Quality Prediction.}} The experimental results on the six datasets are shown in Table~\ref{table:classic} and Table~\ref{table:state-of-arts}, from which we can observe our method achieves the best performance on CSIQ, TID2013, LIVE, IVC, and  LIVE-MultiDist datasets in terms of PLCC, SRCC and KRCC. For KADID-10k, although the pre-trained model DISTS achieves the best results, our method is still comparable without any training procedure introduced.  Moreover, the linear regression results in Fig.~\ref{fig:regression} show that our method presents strong linearity with MOSs, further demonstrating the superior performance of our method.

There are two reasons behind this phenomenon. First, applying WSD as the distance measure for features from VGG16 networks could better align with the visual perception, rooted in the view that the quality is not perceived by direct comparisons. Second, the proposed DeepWSD can deal with different distortion types with satisfied generalization ability and does not rely on a set of hyper-parameters. This is reflected in the challenging KADID-10k and LIVE-MultiDist datasets, on which the promising performance can be also achieved. 

\noindent\textbf{\textit{Quality Optimization.}} 
As indicated in~\cite{ding2021comparison}, a promising way for FR-IQA models evaluation is to use them as objectives for quality optimization. Following this vein, in Fig.~\ref{fig:opt}, we attempt to reconstruct the reference image from the noise image and the white image using different quality measures. This procedure can be formulated as $arg\min _P\;D\left( P,Q \right)$, where $D(P,Q)$ is the FR-IQA model. Moreover, $Q$ is the reference image and $P$ is the reconstructed image initialized by random noise and a pure white image. We compare our optimization results with the two latest deep IQA models, LPIPS and DISTS.  From Fig.~\ref{fig:opt}, the unpleasant artifacts are usually introduced in the optimization results of DISTS, revealing the fidelity may not be well maintained. The LPIPS also presents low-quality results, especially in the rich-texture regions, and cannot fully reconstruct the Lena image when the initial image is the white image. Compared with both DISTS and LPIPS, our method reconstructs the reference image with both high quality and high fidelity, demonstrating the WSD is able to measure the feature distance in a more perceptual way.

\renewcommand{\arraystretch}{1.1}
\begin{table*}[t]
  \centering
  \fontsize{8.6}{8.6}\selectfont
  \begin{threeparttable}
  \caption{Comparison results with different patch sizes and different network structures. The DeepWSD \textit{w}/P4,   DeepWSD \textit{w}/P8 and  DeepWSD \textit{w}/P16 represent the methods with patch sizes set to 4,8,16, respectively. The DeepWSD \textit{w}/VGG16, DeepWSD \textit{w}/AlexNet and DeepWSD \textit{w}/SqueezeNet represent the methods using corresponding network structures. Bold and underline denote the first and second highest scores, respectively.}
  \label{table:patch_size}
    \begin{tabular}{lcccccccccccc}
    \toprule[2pt] \midrule[1pt]
    \multirow{2}{*}{Method} &
    \multicolumn{3}{c}{TID2013~\cite{TID2013}} & \multicolumn{3}{c}{LIVE~\cite{LIVE}} & \multicolumn{3}{c}{CSIQ~\cite{MAD}} &\multicolumn{3}{c}{KADID-10k~\cite{KADID-10k}} \cr
    \cmidrule(lr){2-4} \cmidrule(lr){5-7} \cmidrule(lr){8-10} \cmidrule(lr){11-13}
    &PLCC&SRCC&KRCC&PLCC&SRCC&KRCC&PLCC&SRCC&KRCC&PLCC&SRCC&KRCC \cr
    \hline
    DeepWSD \textit{w}/P4 & \textbf{0.8697} &\textbf{0.8741} &\textbf{0.6932}&\textbf{0.9609} &\textbf{0.9624} &\textbf{0.8378} &\textbf{0.9503} &\textbf{0.9652} &\textbf{0.8297} &\textbf{0.8434} & \textbf{0.8883} &\textbf{0.7076} \\
    DeepWSD \textit{w}/P8 & \underline{0.8512} & \underline{0.8662} & \underline{0.6749} & \underline{0.9586} & \underline{0.9546} & \underline{0.8237} & \underline{0.9379} & \underline{0.9521} & \underline{0.8232} &\underline{0.8154} & \underline{0.8678} & \underline{0.6887} \\
    DeepWSD \textit{w}/P16 & {0.8323} & {0.8489} & {0.6577} &0.9471 &0.9448 &0.8135 &0.9246 &0.9503 &0.8189 & {0.7931} & {0.8423} & {0.6703}\\
    \hline
    DeepWSD \textit{w}/VGG16 &\textbf{0.8697} &\textbf{0.8741} &\textbf{0.6932} &\textbf{0.9609} &\textbf{0.9624} &\textbf{0.8378} &\textbf{0.9503} &\textbf{0.9652} &\textbf{0.8297} &\textbf{0.8434} &\textbf{0.8883} & \textbf{0.7076}\\
    DeepWSD \textit{w}/AlexNet &\underline{0.8471} &0.8266 &0.6479 &0.9237 &0.9518 &0.8175 &\underline{0.9478} &0.9515 &0.7952 &0.7825 &0.8094 &0.6076\\
    DeepWSD \textit{w}/SqueezeNet &0.8354 &\underline{0.8673} &\underline{0.6855} &\underline{0.9536} &\underline{0.9567} &\underline{0.8321} &0.9381 &\underline{0.9597} &\underline{0.8127} &\underline{0.8393} &\underline{0.8801} &\underline{0.7027} \\
    \midrule[1pt] \bottomrule[2pt]
    \end{tabular}
    \end{threeparttable}
\end{table*}

\subsection{Ablation Studies}
\begin{figure}[t]
\centering
      \includegraphics[scale=0.32]{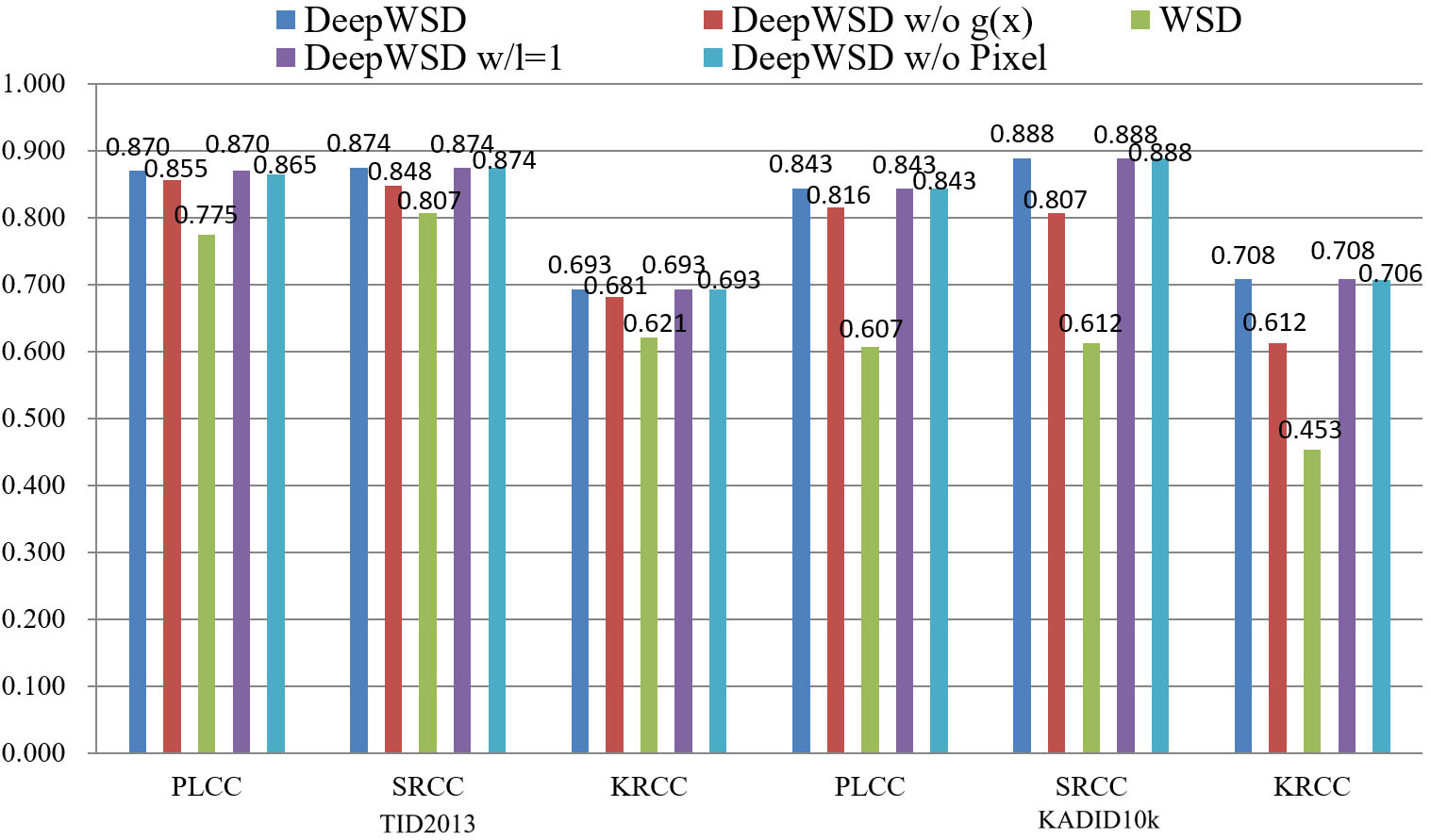}
      \caption{Ablation study results of DeepWSD, DeepWSD \textit{w/o} $g(s)$, WSD and DeepWSD \textit{w/o} Pixel on TID2013 and KADID-10k datasets.}
      \label{fig:g(x)&WSD}
\end{figure}
\balance
\textbf{\textit{Influence of Different Patch Sizes.}} To explore the best patch size for WSD performing on VGG layers, we set various patch sizes in our proposed method and present the comparison results on TID2013 and KADID-10k datasets in Table~\ref{table:patch_size}. We denote the model with patch size set to 4, 8, 16 as DeepWSD \textit{w}/P4, DeepWSD \textit{w}/P8 and DeepWSD \textit{w}/P16, respectively. From the table, we can observe that the best performance is achieved by setting the patch size as 4. The reason lies in that the smaller patch size could benefit the capture of tiny local distortion.

\noindent\textbf{\textit{Ablation of Adaptive Weight  $g(x)$.}} In Eqn.~(\ref{Eq.DeepWSD}), the adaptive wight $g(x)$ is adopted for the Euclidean distance measuring term in the pixel domain. To verify the effectiveness of $g(x)$, we ablate it from Eqn.~(\ref{Eq.DeepWSD}), leading to the model denoted as  DeepWSD \textit{w/o} $g(x)$. The comparison results on TID2013 and KADID-10k datasets are shown in Fig.~\ref{fig:g(x)&WSD}, in which we observe the performance drop significantly in terms of both SRCC and PLCC. This phenomenon demonstrates that the adaptive wight involved in fidelity measurement plays an important role in pursuing perceptual consistency.

\noindent\textbf{\textit{WSD Performed in Pixel Domain.}} 
As we discussed before, the WSD performed in deep feature space can capture the pixel-correlation in a quality-aware manner. To verify such an assumption, we incorporate the WSD in the pixel domain \textit{i.e.,} we measure the WSD of the reference image and distorted image patch-by-patch in the RGB space for image quality estimation. The comparison results on TID2013 and KADID-10k datasets are shown in Fig.~\ref{fig:g(x)&WSD}. We could observe the superior performance our DeepWSD can achieve, demonstrating that deep features provide more appropriate feasible space for quality measuring. Although conducting the WSD directly on the pixel domain is not optimal, we find it can be complementary to the feature domain WSD. As shown in Fig.~\ref{fig:g(x)&WSD}, when we ablate the pixel information from the DeepWSD, the results (denoted as DeepWSD \textit{w/o} Pixel) show the performance drop in terms of PLCC (0.865 \textit{v.s} 0.875) on the TID2013 dataset and KRCC (0.708 \textit{v.s} 0.706)  on the KADID-10k dataset.

\noindent\textbf{\textit{Different Network Structures.}}
 To explore the performance of WSD on different feature encoders, we use the same philosophy to build WSD on the structure of AlexNet~\cite{AlexNet} and SqueezeNet~\cite{SqueezeNet}. The results are shown in TABLE~\ref{table:patch_size}. We observe that the VGG16 backbone can achieve the best performance on different datasets, revealing its effectiveness and high generalization capability. Our observation is also in line with the results of LPIPS~\cite{LPIPS}.

\section{Conclusions}
In this paper, we have developed a new quality measure DeepWSD and shown that simplifying comparing the deep features from the perspective of the probability distribution is effective for quality assessment. The proposed quality measure does not rely on plentiful hyper-parameters and is constructed with the pre-trained network without tuning with quality labels. Experimental results show that the proposed method is well correlated with the subjective evaluations of image quality, based upon the conventional distortions as well as the distortions that typically occur in real-application scenarios. 

The message we are trying to send here is not that the traditional FR-IQA measure with direct feature comparisons should be abandoned. Rather, it is highly expected that the design of FR-IQA should also assimilate the idea of NR-IQA methods, the majority of which are based upon statistical modeling. In particular, the adoption of the WSD that serves as the metric on probability distributions enables the use of a rich groundwork of distribution based methods in FR-IQA. 

\section{Acknowledgments}
This work was supported in part by the Key Project of Science and Technology Innovation 2030 supported by the Ministry of Science and Technology of China (Grant No. 2018AAA0101301), the National Natural Science Foundation of China Grant 61672443, 62022002 and 62176027, in part by Hong Kong GRF - RGC General Research Fund 9042816 (CityU 11209819) and 9042958 (CityU 11203820). Also, this work is supported by the Hong Kong Innovation and Technology Commission (InnoHK Project CIMDA).


\bibliographystyle{ACM-Reference-Format}
\bibliography{sample-base}

\end{document}